# Superperiods and quantum statistics of Laughlin quasiparticles

V. J. Goldman

*Department of Physics, Stony Brook University, Stony Brook, New York 11794-3800, USA*

Superperiodic conductance oscillations were recently observed in the quasiparticle interferometer, where an edge channel of the 1/3 fractional quantum Hall fluid encircles an island of the 2/5 fluid. We present a microscopic model of the origin of the $5h/e$ flux superperiod based on the Haldane-Halperin fractional-statistics hierarchical construction of the 2/5 condensate. Since variation of the applied magnetic field does not affect the charge state of the island, the fundamental period comprises the minimal 2/5 island *neutral* reconstruction. The period consists of incrementing by one the state number of the $e/3$ Laughlin quasielectron circling the island and the concurrent excitation of ten $e/5$ quasiparticles out of the island 2/5 condensate. The Berry phase quantization condition yields anyonic quasiparticle braiding statistics consistent with the hierarchical construction. We further discuss a composite fermion representation of quasiparticles consistent with the superperiods. It is shown to be in one-to-one correspondence with the Haldane-Halperin theory, provided a literal interpretation of the 2/5 condensate as comprised of an integer multiple of two-composite fermion, one-vortex blocks is postulated.

## I. INTRODUCTION

In two spatial dimensions the laws of physics allow existence of particles with fractional exchange statistics, dubbed *anyons*.[1,2] The particles have quantum statistics $\Theta$ if upon exchange the two-particle wave function acquires a phase factor of $\exp(i\pi\Theta)$, and, upon a closed loop, a factor of $\exp(i2\pi\Theta)$. The integer values $\Theta_B = 2j$ and $\Theta_F = 2j+1$ describe the familiar Bose and Fermi exchange statistics, respectively. Upon execution of a closed loop both bosons and fermions produce a phase factor of $+1$, which is unobservable, and the statistical contribution can be safely neglected for the integer statistics particles when describing an interference experiment, such as the Aharonov-Bohm effect. The elementary charged excitations, Laughlin quasiparticles,[3] of a fractional quantum Hall (FQH) electron fluid[4,3] have a fractional electric charge[3,5] and were predicted to be anyons, that is to obey fractional exchange statistics.[6,7]

Recent experiments on electron interferometer devices in the quantum Hall regime, where electrons encircle a two-dimensional (2D) electron island, have reported observation of an Aharonov-Bohm superperiod, implying fractional statistics of Laughlin quasiparticles.[8,9] Experimental results clearly show interference of Laughlin quasiparticles in an edge channel of the filling $f = 1/3$ fluid circling an $f = 2/5$ island. Experimental tests establish: (i) the transport current displaying the interference signal is carried by the $e/3$ Laughlin quasiparticles in 1/3 FQH fluid, see Figs. 4, 8 in Ref. 8, Fig. 1(b) in Ref. 10; (ii) the interference signal has $\Delta_\Phi = 5h/e$ magnetic flux period and the corresponding $\Delta_Q = 2e$ electric charge period, see Ref. 11; and (iii) these superperiods originate in an FQH filling 2/5 island, see Fig. 4(b) in Ref. 9.

The specific model situation proposed here is that of a $-e/3$ quasielectron, which carries the transport current modulated by the interference, encircling an island of the 2/5 FQH fluid. This is the relevant fundamental model since the superperiodic conductance oscillations are observed in the limit of small temperature and excitation, so that the experiment detects superperiodic ground state reconstruction of a 2/5 island embedded in the 1/3 fluid. We develop a microscopic Haldane-Halperin[6,12] hierarchical model consistent with the experimental observations, including the $e/3$ quasiparticle statistics $\Theta_{1/3} = 2/3$. The model is based on the fact that the island is large, containing ~2,000 electrons, so that the fundamental physical processes in the 2/5 island must closely mimic what occurs in the ground state of an infinitely large 2/5 fluid, with an additional constraint that the island is surrounded by the 1/3 fluid. The requirement that the 2/5 island remains in the FQH quasiparticle-containing ground state implies that Coulomb energy must be minimized, even after many periodic island reconstructions.

We first review the physics of quasiparticle-containing ground state of a FQH fluid at a filling factor deviating from the exact filling. Then, we explore the constraints imposed by the topological order[13] incorporated in the Laughlin-Haldane-Halperin FQH theory and show that they lead to superperiodic behavior in the island geometry. The chief effect of the topological order important to the problem under consideration follows from the requirement for an isolated FQH fluid to contain an integer number of electrons. It is easy to see that an island of Haldane-Halperin exact filling condensate at $f = p/(2p+1)$ must contain an integer multiple of $p$ electrons if no quasiparticles are present. The intricacy is to extend this condition to the whole FQH fluid, containing both the condensate and the quasiparticles, incorporating the hierarchy structure of the daughter condensate. Yet this is a manifestation of the FQH topological order. The topological order of the FQH condensates can be parameterized via an anyonic quasiparticle statistical contribution in the Berry phase quantization.

We also show that the quasiparticle construction in terms of composite fermions can be done in one-to-one correspondence. However, quasiparticles are collective excitations of a FQH condensate and thus can not be represented by single composite fermions; mere counting of composite fermions is not sufficient to reproduce the experimental results. We show that the topological order of the microscopic wave function can be recovered, to the extent that the superperiodic behavior is concerned, by postulating that an island of FQH fluid at $f = p/(2p+1)$ must contain an integer multiple of the corresponding $p$-composite fermion condensate blocks, as is the case explicitly in the Laughlin-Haldane-Halperin hierarchy theory.

## II. THE FQH GROUND STATE

We first recall the basic physics of the ground state of an infinitely large 2D electron system in spatially uniform normal magnetic field $B = |\partial_\mathbf{r} \times \mathbf{A}|$ in the quantum limit, assuming spin polarization.[14,15] For $N$ 2D electrons of charge $-e$ embedded in a material with dielectric constant $\varepsilon$ the Hamiltonian is

$$\mathsf{H} = \frac{1}{2\mu}\sum_{j}^{N}[-i\hbar\partial_{\mathbf{r}_j} + e\mathbf{A}(\mathbf{r}_j)]^2 + \frac{1}{4\pi\varepsilon\varepsilon_0}\sum_{j<k}^{N}\frac{e^2}{|\mathbf{r}_j - \mathbf{r}_k|} - \sum_{j}^{N}eV(\mathbf{r}_j), \qquad (1)$$

where $\mathbf{r}_j$ is the $j$-th electron position and $V(\mathbf{r})$ is the potential created in the 2D plane by the positive neutralizing background (that is, the ionized donors and gate electrodes in semiconductor heterostructures).

The non-interacting problem has been solved by Landau.[16] In the symmetric gauge $\mathbf{A}(\mathbf{r}) = \frac{1}{2}\mathbf{B}\times\mathbf{r}$, convenient for $N$ electrons on a disc, the solutions[17] form highly degenerate Landau levels with energies (neglecting spin) $E_n = (\hbar eB/\mu)(n + \frac{1}{2})$, where $n = 0, 1, \cdots$. The ground state wave functions can be written as completely antisymmetric Slater determinants of the basis orbitals $\psi_m^n(\mathbf{r})$; for example, the lowest Landau level $n = 0$ orbitals are

$$\psi_m^0(r,\varphi) = r^m \exp(im\varphi - r^2/4)/\sqrt{2\pi\, 2^m m!}, \tag{2}$$

where $\varphi$ is the azimuthal angle, distances are in units of the magnetic length $\ell_0 = (\hbar/eB)^{1/2}$, and the orbital quantum number $m = 0, 1, \cdots$. Every spin-polarized Landau level has one electron state per Landau quantization area $S_0$, which encloses $h/e$ of magnetic flux, $S_0 = 2\pi\ell_0^2 = h/eB$. The $N$-electron Slater determinant for $p = 1, 2, \cdots$ completely filled Landau levels is

$$\Psi_{p,N} = \begin{Vmatrix} \psi_0^0(z_1) & \psi_0^0(z_2) & \cdots & \psi_0^0(z_N) \\ \psi_1^0(z_1) & \psi_1^0(z_2) & \cdots & \psi_1^0(z_N) \\ \vdots & \vdots & \cdots & \vdots \\ \psi_{N/p-1}^{p-1}(z_1) & \psi_{N/p-1}^{p-1}(z_2) & \cdots & \psi_{N/p-1}^{p-1}(z_N) \end{Vmatrix}. \tag{3}$$

Note that $N$ must be an integer multiple of $p$ in order to have an electron fluid with $p$ completely filled Landau levels.

The ground state solutions of the full Hamiltonian Eq. (1) are many-electron wave functions $\Psi(\mathbf{r}_j)$, which are not known explicitly except in small-system numerical diagonalization. Certain properties of $\Psi(\mathbf{r}_j)$ imposed by the physics of the problem can be reasoned without actually solving the Schrödinger equation with the Hamiltonian Eq. (1). For example, $\Psi(\mathbf{r}_j)$ must be completely antisymmetric under exchange of any two electrons. In order to minimize the Coulomb interaction, $\Psi(\mathbf{r}_j)$ goes to zero as any two electrons approach each other. Also, in the classical limit minimization of the Coulomb energy results in the Poisson equation for the electron charge density $\rho = -e\Psi^*\Psi$, with small quantum corrections, unless a phase transition to a Wigner crystal occurs,[14] resulting in a constant electron density in the interior of the disc.

A spatially uniform electron density $n$ minimizes the energy of interaction with the uniform positively charged neutralizing background. In analogy to the formulation of the BCS theory for a quasiparticle-containing superconductor,[18] a quantum Hall fluid at Landau level filling $\nu = nh/eB$ is considered to consist of an exact filling $f$ incompressible quantum Hall condensate and the charged elementary "excitations" of this condensate, the quasiparticles. While the filing factor $\nu$ is a variable, the quantum Hall exact filling $f$ is a quantum number defined as the value of the *quantized* Hall conductance $\sigma_{XY}$ in units of $e^2/h$ (that is, $f \equiv \sigma_{XY} h/e^2$). Thus, the quantum Hall fluid is conceptually separated into the exact filling condensate and its quasiparticles: $n_f = f\, eB/h$ out of the total electron density goes to form the condensate, and $n_f^{QP} = n - n_f$ goes to form the quasiparticles. This can be expressed as charge conservation of the mean charge density,

$$\rho_\nu = -e[n_f + (n - n_f)] = \rho_f + \rho_f^{QP}. \tag{4}$$

The integer quantized Hall electron fluid ground state condensate is $\Psi_{p,N}$, formed when $f$ is an integer $p$. In the non-interacting problem $\Psi_{p,N}$ is given by Eq. (3). The quasielectron elementary excitation is an electron in the $p+1$ Landau level, the quasihole is an absence of an electron in the otherwise filled $p$-th Landau level. A ground state containing quasielectrons has filling factor $\nu > f$, and a ground state containing quasiholes has $\nu < f$. It can be argued that the interacting integer quantum Hall problem can be mapped onto the non-interacting problem: one can not speak of $p$ completely occupied Landau levels in the interacting problem, still the quantum Hall gap does form at filling $f = p$, and the filling factor $\nu$ can be varied away from the exact filling, remaining on the $f = p$ plateau, to quasiparticle-containing ground states.

Laughlin has shown[3] that certain kinds of trial wave functions capture the essential physics of the highly-correlated FQH ground states, and, in particular, the Laughlin wave functions are known to be exact solutions for short-range interaction (Haldane $V_1$ pseudopotential)[15] Hamiltonians. The filling $f = 1/3$ Laughlin wave function[3,14] for $N$ electrons is

$$\Psi_{1/3} = \prod_{j<k}^{N} (z_j - z_k)^3 \times \exp(-\tfrac{1}{4} \sum_{l}^{N} |z_l|^2), \qquad (5)$$

where $z_j = r_j e^{i\varphi_j}$ are the complex electron coordinates on the disc. Like the integer quantum Hall fluid of interacting electrons, the FQH incompressible condensate has spatially constant charge density $\rho_f$, the compressibility of the total electron system, which allows $\nu$ to deviate from $f$, results from variation of the quasiparticle mean density $\rho_f^{QP}$, Eq. (4).

The two kinds of FQH quasiparticles are the quasiholes, quantized vortices (deficiencies) in the condensate, created in the FQH fluid ground state for $\nu < f$, and the quasielectrons, quantized excesses in the local condensate density, created for $\nu > f$. Thus, a FQH fluid is comprised of an exact filling condensate and the quasiparticles, which are "on top" or "in addition" to the condensate. The local 2D electron density fluctuates from the flat condensate density in a several magnetic length vicinity of the quasiparticle position. For a dilute quasiparticle density ($\nu \cong f$), integrating the 2D electron density over sufficiently large area gives a quantized deviation from what is obtained by integrating the condensate density only, this is the quasiparticle charge.

The above can be illustrated by the following example. A FQH fluid state having a quasihole localized at $\xi_\alpha$ is given by (not normalized) wave function[3]

$$\Psi_{1/3}^{+\xi_\alpha} = \prod_{n=1}^{N} (z_n - \xi_\alpha) \Psi_{1/3}. \qquad (6)$$

The quasiparticle-containing electron fluid described by $\Psi_{1/3}^{+\xi_\alpha}$ still contains $N$ electrons, and it can be conceptually separated into the $f = 1/3$ exact filling FQH condensate, containing charge $-e(N+1/3)$, and the quasihole, charge $+e/3$. The condensate charge density,

$$\rho_{1/3} = (-e/\Sigma) \int_\Sigma dS \, (\Psi_{1/3})^* \Psi_{1/3} = -e(eB/3h), \qquad (7)$$

determined only by the magnetic field, is constant in the interior of the disc. The integral is over area $\Sigma$ well within the disc boundary, and assumes normalized wave function. Thus, in a given field the FQH condensate is incompressible. The quasihole charge distribution is localized on the length scale of several $\ell_0$; it oscillates, but the quasihole charge $q$, obtained by integrating over a sufficiently large area $\Sigma$ centered on $\xi_\alpha$, well within the disc boundary,

$$q = -e \int_\Sigma dS \, (\Psi_{1/3}^{+\xi_\alpha})^* \Psi_{1/3}^{+\xi_\alpha} - \rho_{1/3} \Sigma \qquad (8)$$

is quantized to $+e/3$. In the thermodynamic limit $N \to \infty$, when $\Sigma$ can be large enough to contain many quasiholes, we can analogously define mean quasihole charge density $\rho_{1/3}^{QH}$, average over $\Sigma$. The total FQH fluid is incompressible only in the sense that it is gapped, that is, it takes a finite energy to create a quasiparticle when $\nu$ is varied.

One can use charge conservation to calculate various FQH ground state properties. If we consider an FQH fluid at filling $\nu < f$, the condensate charge density is $\rho_f = -fe^2B/h$, and the quasihole mean charge density is $\rho_f^{QH} = (f-\nu)e^2B/h$. These are convenient to write as

$$\rho_f = -fe/S_0 \tag{9}$$

and

$$\rho_f^{QH} = (f-\nu)e/S_0, \tag{10}$$

where the Landau quantization area $S_0$ encloses $h/e$ of flux, $S_0 = 2\pi\ell_0^2 = h/eB$. An arbitrary area $S$ contains electronic charge $\rho_\nu S = -\nu eS/S_0$, split between the condensate $\rho_f S = -feS/S_0$ and the quasihole charge $\rho_f^{QH} S = (f-\nu)eS/S_0$. Neither the expectation value of the number of electrons in area $S$, $\langle N_e \rangle = \int_S dS \, \Psi^*\Psi = \nu S/S_0$, nor the number of quasiparticles, $\langle N_f^{QP} \rangle = (e/q)(f-\nu)S/S_0$, must be integer since area $S$ is arbitrary, is not defined by a physical constraint. This also applies to the number of electrons in the condensate $\langle N_f \rangle = fS/S_0 = \langle N_e \rangle - (q/e)\langle N_f^{QP} \rangle$.

When magnetic field is varied adiabatically, the ground state electron density $\rho_\nu$ is not affected because it neutralizes the positive background, minimizing Coulomb energy. Any necessary relaxation of the FQH fluid to the ground state can be accomplished via a small, but finite dissipative diagonal conductivity $\sigma_{XX}$. Thus, in the ground state, quasiparticles are excited out of the FQH condensate. For example, if magnetic field is increased so that a fixed area $S$ contains one more $h/e$ of flux (one more $S_0$ fits into $S$), condensate $\rho_f$ increases by $-ef/S$, quasihole $\rho_f^{QH}$ increases by $+ef/S$, and the sum, the electronic $\rho_\nu$ is not changed. Thus $\langle N_e \rangle$ is not changed, $\langle N_f \rangle$ increases by $f$, and $\langle N_f^{QH} \rangle$ increases by $(e/q)f$ [or, equivalently, $\langle N_f^{QE} \rangle$ decreases by $(|e/q|)f$].

So far we discussed mean, average particle densities. Clearly, at the microscopic level the quantized quasiparticles are excited out of condensate in steps of one. As we will see below, in a constricted geometry condensate reconstruction occurs in quantized steps also. These steps correspond to one quasihole for the $f = 1/(2j+1)$ primary Laughlin condensates only. The specific minimal steps of an FQH condensate reconstruction are determined by its Haldane-Halperin hierarchical structure, or, equivalently, by its structure in the composite fermion model. At the mean level, the microscopic condensate reconstruction steps must reproduce the FQH fluid filing factor variation described above. Thus, we see that the quasiparticle content in the FQH ground state is not arbitrary, but is uniquely determined by the filling factor and the minimization of the system's Coulomb energy.

### III. HALDANE-HALPERIN HIERARCHY

In this Section we discuss the electron system reconstruction consistent with incompressibility of FQH condensates and minimization of Coulomb energy, without regard to quantization of the encircling quasiparticle orbitals, considered in Section. IV. In the Haldane-Halperin hierarchy theory[6,12] the $f = 2/5$ FQH condensate[19] consists of a "maximum density droplet" (MDD) condensate of $-e/3$ quasielectrons on top of the exact filling 1/3 condensate. This can be written as

$$\Psi_{2/5} = \Psi_{1/3} \otimes \Psi_{MDD}, \tag{11}$$

where $\Psi_{1/3}(z_1,\ldots,z_N)$ is the Laughlin wave function Eq. (5), the Halperin's $-e/3$ quasielectron condensate wave function[6] is

$$\Psi_{MDD}(\xi_1,\ldots,\xi_{N'}) = \prod_{j<k}^{N'}(\xi_j - \xi_k)^2(\xi_j - \xi_k)^{-1/3} \times \exp(-\tfrac{1}{12}\sum_n^{N'}|\xi_n|^2), \tag{12}$$

and $\xi_j$ are the quasielectron coordinates in units of $\ell_0$. The Hilbert space of the wave function Eq. (11) is the product of the $N$ electron and the $N'$ quasielectron spaces.

The density of the MDD $-e/3$ condensate $n_{1/3}^{QE} = 1/5S_0$ is determined by the anyonic statistics of quasielectrons,[6,14] the Landau quantization area $S_0 = 2\pi\ell_0^2 = h/eB$. The resulting total electron charge density $-en$ corresponds to the $f = 2/5$ exact filling condensate:

$$\rho_{2/5} = \frac{-e}{3S_0} + \frac{-e/3}{5S_0} = \frac{-2e}{5S_0}. \tag{13}$$

Since the 2/5 condensate is incompressible, both its hierarchical constituents, the 1/3 condensate and the MDD of $-e/3$ quasielectrons are incompressible. There is one electron per $3S_0$ in $\Psi_{1/3}(z_1,\ldots,z_N)$, the number of quasielectrons $N' = 3N/5$, thus an isolated 2/5 condensate must contain an even number of electrons, $N + N'/3 = (5/3 + 1/3)N' = 2N'$, in order to have an integer number of quasielectrons.

Thus, the 2/5 island embedded in 1/3 FQH fluid is understood as an MDD island of $-e/3$ quasielectrons on top of the 1/3 condensate, the 1/3 condensate extending beyond the MDD island and completely surrounding it, see Fig. 1. The hierarchical wave function can be written as

$$\Psi_{2/5 \text{ in } 1/3} = \Psi_{1/3}(z_1,\ldots,z_N) \otimes \Psi_{MDD}(\xi_1,\ldots,\xi_{N'}), \tag{14}$$

with $N \gg 5N'/3$. The semiclassical island area is $5N'S_0 = 5N'h/eB$, where the magnetic field corresponds to the exact island filling. Like an isolated 2/5 condensate, the 2/5 island must contain an even number of electrons $2N'$. This conclusion holds because 2/5 and 1/3 are the immediate parent-daughter FQH condensates in the hierarchy. Note that this immediate parent-daughter relationship also implies that no other incompressible or compressible electron state exists in between the two.

The elementary charged excitations of the 2/5 condensate are the $\pm e/5$ quasielectrons and quasiholes, excited out of the condensate when the FQH fluid filling $\nu$ deviates from the exact filling, 2/5. The mean density of the $\pm e/5$ quasiparticles $n_{2/5}^{QP}$ can be obtained from conservation of the total electronic charge, Eqs. (4,13),

$$\nu = nS_0 = f \pm (e/5)n_{2/5}^{QP}(S_0/e). \tag{15}$$

This gives $n_{2/5}^{QP} = \pm 5(f-\nu)/S_0$, where quasiholes are excited for $\nu < f$ and quasielectrons for $\nu > f$. In the island geometry, deviation of $\nu$ from $f$ may also cause a change in the number $N_{1/3}^{QE}$ of the MDD $-e/3$ quasielectrons:

$$N_{1/3}^{QE} = n_{1/3}^{QE} S = S/5S_0 \qquad (16)$$

in the island of area $S$. Recalling that the MDD quasielectron condensate is incompressible, the change in their number in a fixed area $S$ can be accomplished only by variation of $S_0$, that is magnetic field.

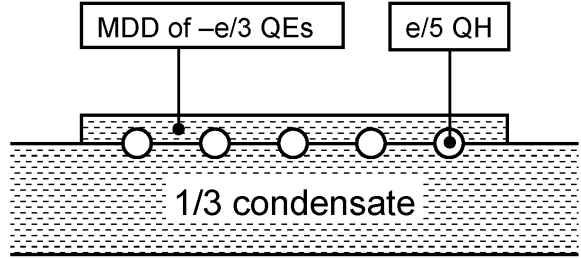

FIG. 1. Illustration of the 2/5 island surrounded by 1/3 fractional quantum Hall fluid in the Haldane-Halperin hierarchy theory. The total 2D electron system is broken into three components: the incompressible 1/3 exact filling FQH condensate, the incompressible maximum density droplet (MDD) of $-e/3$ quasielectrons, and the excited $e/5$ quasiparticles, accommodating the $\nu < 2/5$ situation. The path of the $-e/3$ quasielectron which carries the transport current in the 1/3 fluid encircles the 2/5 island.

Increasing $B$ results in decrease of $S_0 = h/eB$. When the total flux through the island $\Phi = BS$ is increased by $S\delta B = 5h/e$, the number of the MDD quasielectrons $N_{1/3}^{QE} = eBS/5h$ in area $S$ is incremented by one. Concurrently, the $f = 2/5$ island condensate electron density increases by $\delta\rho_{2/5} = -ef\delta B/h = -ef(S\delta B)/Sh = -2e/S$, that is, the total island condensate charge deviates by $-2e$ from neutrality. The mismatched condensate charge is exactly compensated by excitation of ten $e/5$ island quasiholes, Eq. (15), total charge $+2e$, restoring charge neutrality of the total 2D electron system and the fixed positive background. We therefore identify this *minimal* (one MDD quasielectron) microscopic quasiparticle reconstruction of the island with the observed[8,9] $\Delta_\Phi = 5h/e$ flux periodicity. Within the period, increasing $B$, one $-e/3$ quasielectron is added to the MDD because $S_0$ decreases, the $f = 1/3$ condensate charge increases by $-5e/3$ because $\rho_{1/3} = -e/3S_0$ increases, and ten $+e/5$ quasiholes are excited.

The $\pm e/5$ quasiparticles are excitations of the 2/5 condensate at the *next* level of the hierarchy, in contrast to the MDD quasielectrons, which are part of the 2/5 daughter condensate. Thus, excitation of $\pm e/5$ quasiparticles does not affect the $-e/3$ MDD, which is an incompressible constituent of the exact filling 2/5 condensate. To see this clearly, consider an isolated disc of $N_e$ electrons. At $B$ corresponding to exact filling $f = 1/3$, the flux through the disc is $\Phi = 3N_e h/e$, the constant condensate charge density $\rho_{1/3} = -e/3S_0$ is equal to the electron charge density. Increasing $B$ to $B' = B(1 + 1/3N_e)$, so that $\Phi' = \Phi + h/e$, creates one $e/3$ quasihole. The Landau quantization area $S_0$ decreases to $S'_0 = S_0 N_e/(N_e + 1/3)$, so that the exact filling (flat) condensate density $\rho_{1/3} = -e/3S'_0$ increases exactly by $-e/3S$ for the whole disc; the electron density $\rho_v$ develops a notch at the quasihole position, the total electronic charge is still $-eN_e$. Thus, a quantized quasihole is excited out of the 1/3 condensate, the 1/3 condensate charge is quantized in units of $e/3$.

Similarly, an isolated disc of the 2/5 condensate, with no quasiparticles present, by the hierarchy construction Eq. (11) requires an integer number $N'$ of MDD quasielectrons, and must contain an even number $N_e = 2N'$ of electrons and enclose $5N'$ of flux. This also satisfies the additional constraint that an isolated 1/3 FQH fluid, consisting of the 1/3 condensate and its $e/3$ quasiparticles, contain an integer number of electrons. Naively, we can repeat the above argument for one quasiparticle excitation. Increasing magnetic field $B$ to $B' = B(1 + 1/5N_e)$, so that $\Phi' = \Phi + h/2e$, results in excitation of one $e/5$ quasihole. The Landau quantization area $S'_0$ decreases to $S'_0 = S_0 N_e/(N_e + 1/5)$, so that the exact filling condensate density $\rho_{2/5} = -2e/5S_0$ increases exactly by $-e/5S$ for the whole disc, the total electronic charge is still $-eN_e$. Thus, a quasihole is excited out of the 2/5 condensate, condensate density increases, disc area does not change.

However, requiring the $e/5$ quasihole be excitation of the whole 2/5 condensate, 1/6-th of its charge ($e/30$) must come out of the $-e/3$ MDD, and 5/6-th ($e/6$) out of the 1/3 condensate to maintain exactness of the 2/5 condensate filling. Then, the 1/3 condensate and the $-e/3$ MDD condensate would contain non-integer multiples of $e/3$, and the disc area would contain non-integer number of MDD quasielectrons, $N_{1/3}^{QE} = N' + 1/10$. Excitation of two $e/5$ quasiholes maintains the 1/3 condensate as containing an integer multiple of $e/3$, but the MDD component still has non-integer $N_{1/3}^{QE} = N' + 1/5$. This consideration can be extended until ten $e/5$ quasiholes are excited, when both the MDD and the 1/3 condensate components contain integer multiples of $e/3$. Thus the *minimal* isolated 2/5 island reconstruction consistent with the Haldane-Halperin hierarchy involves excitation of ten $e/5$ quasiholes and concurrent increment by one in the number of MDD quasielectrons. This is equivalent to the requirement that the 2/5 condensate contain an even number of electrons, even when $e/5$ quasiparticles are present. In other words, the electronic charge liberated by excitation of $e/5$ quasiholes can be accommodated by the 2/5 island condensate only in increments of $2e$. A transfer of charge between the surrounding 1/3 fluid and the 2/5 island in units of $e/3$, which corresponds to one MDD quasielectron, does not restore the 2/5 island to the correct MDD and the 1/3 condensate composition. Thus, the requirement that the number of MDD quasielectrons be an integer leads to the $\Delta_\Phi = 5h/e$ island reconstruction superperiod.

It can be argued that in the present geometry the $\pm e/5$ quasiparticles may be excited out of the 1/3 condensate hierarchical component of the 2/5 island condensate, leaving the MDD condensate unaffected, because the 1/3 condensate extends to the electron reservoirs and thus its charge is not a sharp observable. Such a process, however, can maintain the correct ratio between the MDD and the 1/3 condensates, but results in net dipole charging of the island. A true period must contain both charging and discharging *sub-periodic* steps, if any. Several alternative sub-periodic 2/5 island reconstructions are possible. Some examples are considered below and in Sect. VII. However, such processes lead to net 2/5 island monopole or dipole charging and, if repeated many times, eventually to huge Coulomb energies, and thus are not viable candidates for a periodic behavior, as discussed in Ref. 8. The precise sequence of sub-periodic island reconstruction steps, minimizing the total energy of the system most likely depends on the details of the confining potential.

The following sequence of ten single-quasiparticle sub-periodic excitation steps should provide un upper bound on the charging energy involved. Starting at the neutral (equilibrium) 2/5 island area $S$ and field $B$, the first step occurs at $B'$ such that area $S$ contains $h/2e$ more flux, adding an $e/5$ quasihole. Since the 2/5 island must contain flux an integer multiple of $5h/e$ (an integer number of MDD quasielectrons), the island area shrinks by ½$S_0$. The original area $S$ now has net charge $e/5 - e/6 = e/30$, in an annulus of uncompensated fixed positive background at the island boundary, neglecting order $eS_0/S$. After fifth of such steps the island has an $e/6$ net annulus charge distribution. Soon thereafter it becomes energetically favorable for the 2/5 island to reconstruct, that is, to acquire an $-e/3$ quasielectron from the 1/3 condensate. This expands its area to include one more MDD quasielectron, that is, $5h/e$ more of flux, and the 2/5 island acquires approximately $-e/6$ net annulus charge, just outside the original $S$. The subsequent $e/5$ quasihole excitation steps discharge the annulus in five steps of $e/30$. The period is completed when the island returns to area $S$ and the net-neutral condition. The maximum macroscopic charging energy during such period is estimated to be under 0.1 K, much less than the FQH gap.

### IV. BERRY PHASE

In the extreme quantum limit of zero temperature and excitation, the interferometer physics can be mapped adiabatically on the resonant tunneling problem,[5,20] the fundamental Berry phase periodicities of the two dynamical processes ought to be the same. The many-electron wave functions of the kind considered in Sect. II contain all the information about the FQH fluids, however, in many situations of interest they are not known explicitly, and numerical solutions are limited to too few electrons. The quasiparticles are collective excitations of the many-electron system, and when there are more than one quasiparticle present it may be difficult to understand what features of the many-electron wave function can be identified with individual quasiparticles. Thus, thinking in terms of a few weakly-interacting (except for the nonlocal statistical interaction) quasiparticles moving on top of the FQH condensate vacuum, instead of fermionic, but strongly-interacting electrons, can greatly simplify description of a FQH system.[7]

In the symmetric gauge $\mathbf{A}(\mathbf{r}) = \frac{1}{2}\mathbf{B} \times \mathbf{r}$ a rotationally-invariant system ground state wave functions $\Psi_M$ are eigenstates of the total angular momentum and can labeled by the quantum number $M$.[14,20] A single particle, including single anyon, Aharonov-Bohm problem can be solved explicitly.[21] The Hamiltonian is

$$\mathsf{H} = \frac{1}{2\mu}\left(-i\hbar\partial_{\mathbf{r}} - q\mathbf{A}\right)^2 + qV(\mathbf{r}), \tag{17}$$

where $V(\mathbf{r})$ is the scalar potential that localizes the particle; adiabatic variation of $V(\mathbf{r},t)$ can be used to thread the particle through the desired path $C$. The solution has the form

$$\Psi(\mathbf{r},t) = \exp[i\frac{q}{\hbar}\Phi(\mathbf{r})]\Psi', \tag{18}$$

where $\Psi'$ satisfies the Schrödinger equation without the vector potential, and the "flux function" path integral

$$\Phi(\mathbf{r}) = \int_O^{\mathbf{r}} \mathbf{A}(\mathbf{r}') \cdot d\mathbf{r}' \tag{19}$$

is measured from the reference point $O$. The closed eigenstate Aharonov-Bohm orbitals satisfy

$$\gamma_M \equiv \frac{q}{\hbar} \oint_C \mathbf{A}(\mathbf{r}) \cdot d\mathbf{r} = \frac{q}{\hbar} \Phi = 2\pi M, \tag{20}$$

where $M$ is an integer. The Aharonov-Bohm phase factor in Eq. (18) is special case of the Berry phase, $\exp(i\gamma)$,

$$\gamma = i \oint_C d\mathfrak{R} \left\langle \Psi(\mathfrak{R},\mathfrak{R}') \left| \frac{\partial}{\partial \mathfrak{R}} \Psi(\mathfrak{R},\mathfrak{R}') \right. \right\rangle \tag{21}$$

and can be obtained by an explicit calculation using potential $V[\mathbf{r} - \mathfrak{R}(t)]$ to localize the particle near position $\mathfrak{R}$ to take it adiabatically via the path $C$ encircling the particle at $\mathfrak{R}'$.[21]

Reference 7 used the adiabatic theorem to calculate the Berry phase of quasiholes in the $f = 1/3$ Laughlin wave function Eqs. (5,6) on a disc. When a quasihole adiabatically executes a closed path the wave function acquires the Berry phase. Taking counterclockwise as the positive direction,[17] they found the difference between an "empty" loop, containing the FQH condensate "vacuum" only, and a loop containing another quasihole to be $\Delta \gamma_{1/3} = 4\pi/3$, identified as the statistical contribution. Generalizing the above, we assert that in a main hierarchy sequence FQH fluid at $f$, containing only one kind of quasiparticles, the quasiparticle orbitals are quantized so that the total Berry phase, combining the Aharonov-Bohm and the statistical contributions, is an integer multiple of $2\pi$,

$$\gamma_M = \frac{q}{\hbar} \Phi + 2\pi \Theta_f N_f = 2\pi M, \tag{22}$$

where $\Theta_f$ is the statistical parameter of the quasiparticles, defined so that upon exchange the wave function acquires a phase factor $\exp(i\pi\Theta)$, and $N_f$ is the number of other quasiparticles within the orbital.

Generalizing further, to include the situation when more than one kind of quasiparticles is present, however specifically for a $q = -e/3$ quasielectron encircling the 2/5 island, we write

$$\gamma_M = \frac{q}{\hbar} \Phi + 2\pi(\Theta_{1/3} N_{1/3} + \Theta_{2/5}^{-1/3} N_{2/5}) = 2\pi M, \tag{23}$$

$\Theta_{1/3} = \Theta_{-1/3}$ is the statistics of $e/3$ quasiparticles, $N_{1/3}$ is the number of the MDD $-e/3$ quasielectrons being encircled, and $\Theta_{2/5}^{-1/3}$ and $N_{2/5}$ refer to the $e/5$ island quasiparticles. The relative statistics $\Theta_{2/5}^{-1/3}$ is defined so that an $-e/3$ quasielectron picks up a statistical phase factor of $\exp(i2\pi\Theta_{2/5}^{-1/3})$ upon execution of a loop around a 2/5 quasihole. Note that the "real" $e/3$ quasiparticles cannot exist in the 2/5 FQH fluid, the MDD quasielectrons are a hierarchical constituent of the 2/5 condensate.

When the chemical potential moves between two successive states of the encircling quasiparticle, $\Psi_M \to \Psi_{M+\Delta_M}$, the change in the phase of the system wave function is $2\pi$:

$$\Delta_\gamma \equiv \gamma_{M+\Delta_M} - \gamma_M = \frac{q}{\hbar} \Delta_\Phi + 2\pi(\Theta_{1/3} \Delta_{N_{1/3}} + \Theta_{2/5}^{-1/3} \Delta_{N_{2/5}}) = 2\pi. \tag{24}$$

Here $\Delta_\Phi$ is the flux period and $\Delta_N$ refer to the corresponding change in the number of the island quasiparticles. Clearly, the $M \to M + \Delta_M$ transition entails $\Delta_{N_{1/3}} = 1$, one more MDD quasielectron (this term is missing in the equations of Refs. 8 and 9). The corresponding increase of the flux through the island is $\Delta_\Phi = 5BS_0 = 5h/e$, see Eq. (16). Concurrently $\Delta_{N_{2/5}} = 10$ of $e/5$ quasiholes are excited in the 2/5 island, as discussed in Sect. III. Upon substitution of these numbers Eq. (24) becomes

$$\frac{\Delta_\gamma}{2\pi} = -\frac{5}{3} + \Theta_{1/3} + 10\,\Theta_{2/5}^{-1/3} = 1. \qquad (25)$$

Two concurrent physical processes comprise the Berry phase period: increase by one of the state number of the encircling $-e/3$ quasielectron because the 2/5 island contains one more MDD quasielectron, and the excitation of ten $e/5$ quasiholes in the island. Thus, the physics of the problem under consideration leads to interpretation of Eq. (25) as two simultaneous equations, each with an integer statistical Berry phase period:

$$1/3 + \Theta_{1/3} = 1, \text{ and} \qquad (26a)$$

$$10\,\Theta_{2/5}^{-1/3} = 2. \qquad (26b)$$

Eq. (26a) is identical to that in quantum antidots on the $f = 1/3$ plateau.[5,20] Eq. (26b) can be understood as sum of two $5\,\Theta_{2/5}^{-1/3} = 1$ equations, each for the expected two kinds of the $e/5$ quasiparticle excitation of the $f = 2/5$ condensate. These are solved by

$$\Theta_{1/3} = 2/3, \text{ and} \qquad (27a)$$

$$\Theta_{2/5}^{-1/3} = 1/5. \qquad (27b)$$

The value $\Theta_{1/3} = 2/3$ is in agreement with the expectation,[6,7,22] and with the quantum antidot experiments.[5,20,23] The value $\Theta_{2/5}^{-1/3} = 1/5$ appears to be consistent with what would be obtained in a Berry phase calculation similar to that of Ref. 7, by virtue of the Cauchy's theorem, treating the 2/5 island as the MDD of the $-e/3$ quasielectrons, and including the charge deficiency in the 2/5 condensate created by excitation of an $e/5$ quasihole vortex, and maintaining the path of the adiabatically encircling $-e/3$ quasielectron fixed. Also, note that a $2.5h/e$ period (excitation of five island quasiparticles) were possible if $\Theta_{1/3}$ were an integer; thus the observed superperiod entails both $\Theta_{2/5}^{-1/3}$ and $\Theta_{1/3}$ must be anyonic. The relative (mutual) statistics of quasiparticles of the two FQH condensates at different filling are meaningful because both quasiparticle kinds are different collective excitations of a single highly correlated electron system comprising the parent-daughter FQH fluid with different fillings. The topological order of FQH condensates[13] is thus manifested by the anyonic statistics of their quasiparticles.

## V. EDGE CHANNEL STRUCTURE

The common flux period of $\Delta_\Phi = 5h/e$ results from the fundamental coincidence of the parent-daughter relation between the two FQH condensates and only one area appearing in the path integral in the model of Sect. III, IV. The experiment probes the reconstruction of the 2/5 island embedded in 1/3 fluid by measuring oscillating conductance of the interferometer. Thus, though not directly relevant to the origin of the superperiodic reconstruction of the island ground state, it is interesting to consider a dynamical transport model where such conductance oscillations may arise. The edge channel structure model consistent with the above microscopic model can be envisioned as an incompressible 1/3 edge ring enclosing the 2/5 island.[10,11] The 2/5 island may contain $e/5$ quasiparticles in the ground state. Since 1/3 and 2/5 FQH fluids are the immediate parent-daughter hierarchical states, there is no other compressible or incompressible FQH state in between.

The transport current is carried into the interferometer region by the 1/3 edge channel fluid extending beyond the two constrictions. We recall that the transport current is the difference of currents in the two counter-propagating incompressible edge channels held at different potentials.[14] The quantum Hall filling of the current-carrying channels is determined by the constriction saddle-point electron density, similar to the $f = 1$ edge ring in the integer regime. In the integer regime the island interior contains a compressible disc with $\nu > 1$, thus accommodating the ~20% higher electron density near the island center. In the fractional regime the 2/5 island forms in the higher electron density region.

In both regimes, no transport current is carried through the island interior because the equipotentials of the confining potential are closed in the interior. No equipotential extends between the opposite edges in the island, see Fig. 2(b) of Ref. 11. The confining potential limits the radial width of the 1/3 edge channel ring to about $t \approx 115$ nm,[8,9,11] about $15\ell_0$. The tunnel coupling over this distance is quite considerable, $\sim \exp[-(t/2\sqrt{3\pi}\ell_0)^2] \sim 2 \times 10^{-3}$, so that the encircling $-e/3$ quasielectrons quantum-coherently extend to the 2/5 island, which they cannot penetrate. The resulting chiral Aharonov-Bohm path effectively skirts the higher filling island. Thus the only area is defined by the encircling quasielectrons' Aharonov-Bohm path skirting the 2/5 island circumference, and the $\Delta_\Phi = 5h/e$ periodicity is robust under moderate island perturbations, such as application of a front gate voltage.[11] This model is different from a fortuitous coincidence of two areas and only one type of quasiparticles in the model of Ref. 24.

## VI. COMPOSITE FERMIONS

In the composite fermion (CF) model,[25] the FQH condensates at filling $f = p/(2jp+1)$ are modeled as integer quantum Hall states of the $2j$–vortex CFs at filling $p$. Here, $p, j = 1, 2, \ldots$. The unprojected CF wave functions are constructed as the product

$$\Psi^{CF}_{p/(2jp+1)} = D^{2j}\Psi_{p,N} \tag{28}$$

of a Jastrow factor

$$D^{2j} = \prod_{l<k}^{N}(z_l - z_k)^{2j} \tag{29}$$

and the Slater determinants $\Psi_{p,N}$ for $p$ completely filled Landau levels, each with $N/p$ electrons, Eq. (3).

The vortex attachment, formally implemented as a Chern-Simons[26] singular gauge transformation $\mathbf{A} \rightarrow \mathbf{A} + \mathbf{a}$, can transmute 2D electrons into composite anyons,[2,27] composite bosons,[28-31] and composite fermions.[32,33] The Chern-Simons gauge field

$$\mathbf{a}(\mathbf{r}) = \frac{\Phi^*}{2\pi}\sum_l \partial_\mathbf{r}\varphi_l = \frac{\Phi^*}{2\pi}\sum_l \frac{\hat{\mathbf{z}} \times (\mathbf{r} - \mathbf{r}_l)}{|\mathbf{r} - \mathbf{r}_l|^2}, \tag{30}$$

where $\hat{\mathbf{z}} \parallel \mathbf{B}$, attaches fictitious flux $\Phi^*$ to each electron at $\mathbf{r}_l$. To obtain CFs, we bind $\Phi^* = -2jh/e$ of Chern-Simons flux to each 2D electron, the minus sign here means that the fictitious flux opposes the physical applied $B$. The "flux quantization" in units of $h/e$ is not physical, but is imposed here by the desire to obtain composite particles with integer, not anyonic exchange statistics.[2] In the mean field approximation, neglecting quantum field fluctuations, for a spatially uniform 2D electron number density $n$ the vortex attachment "absorbs" $2jnh/e$ out of the physical $\Phi$, so that CFs experience effective mean field $B^{CF} = B - 2jnh/e$. Defining effective CF "Landau level" filling $\nu^{CF} = nh/eB^{CF}$, the integer quantum Hall effect of CFs occurs when $\nu^{CF} = p$, an integer (by construction, CF density is equal to the electron density $n$).

The FQH condensates at $f = p/(2jp+1)$ can be represented as a collection of CF building blocks, each consisting of $p$ of $2j$-vortex CFs and an additional vortex, Fig. 2. The CF condensate block has 2D charge density $\rho_f = -ep/(2jp+1)S_0 = -en$ and occupies area $(2jp+1)S_0$. It may be tempting to think of the quasiparticle elementary excitations as a condensate block with an added (quasielectron) or subtracted (quasihole) $2j$-vortex CF, replacing (annihilating) the parent FQH condensate block. However, the quasiparticles are collective excitations of the parent FQH condensate, and should not be thought of as literally a small CF-containing building block with a well-defined area and charge density. Another way of saying this is that a particular microscopic non-interacting electron Landau level distribution is washed out upon multiplication by the Jastrow factor, Eq. (29), and does not represent the resulting microscopic CF configuration, like the interacting electron integer quantum Hall condensate is not given by Eq. (3). It is well known that the actual quasiparticle charge density oscillates over a large area $\gg (2jp+1)S_0$, while the characteristic magnitude of the charge deficiency or excess is small compared to $\rho_f$. Considering only the CF content of the resulting FQH electron fluid configuration as in Fig. 2, without the additional constraints imposed by the (condensate plus quasiparticles) structure of total fluid, neglects the topological order and incompressibility of the FQH condensates.

Focusing on the $j=1$ two-vortex CFs, the CF model building-block equivalents of the Laughlin quasiparticles are shown in Fig. 2(b). It is easy to see that there is a one-to-one correspondence between the CF model and the Laughlin-Haldane-Halperin construction at this level, provided the CF building blocks are interpreted as the minimal units. In particular, the hierarchy requirement that an isolated FQH exact-filling (no quasiparticles present) condensate at $f = p/(2p+1)$ contain an integer multiple of $p$ electrons and enclose the same integer $N/p$ times $(2p+1)$ of flux is equivalent to the condensate containing an integer multiple of CF blocks, each block comprised of $p$ electrons and $(2p+1)$ of flux. Increasing $B$ so that $h/e$ of flux is added to such isolated FQH condensate excites $p$ quasiholes, one in each of CF "Landau levels". The resulting $(p-1)$ of $f = p/(2p+1)$ condensate blocks and the one added vortex are re-assembled into $p$ of $(p-1)/(2p-1)$ quasihole blocks. Reducing $B$ so that $h/e$ of flux is subtracted from such isolated FQH condensate excites $p$ quasielectrons, one in each of CF "Landau levels"; $(p+1)$ of $f = p/(2p+1)$ condensate blocks and the one subtracted vortex are re-assembled into $p$ of $(p+1)/(2p+3)$ quasielectron blocks. We stress again that such literal interpretation of the Laughlin quasiparticles as small CF blocks neglects the topological order and incompressibility of the parent FQH condensate and thus leads to incorrect predictions.

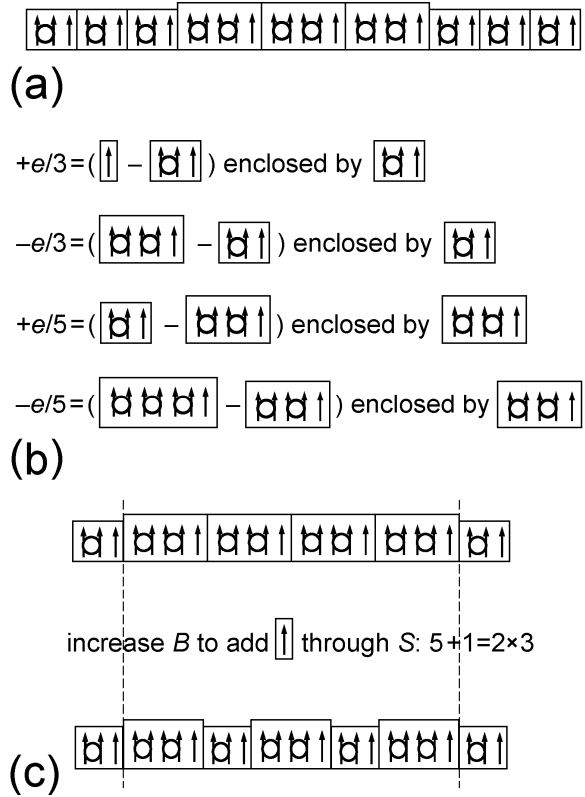

FIG. 2. Composite fermion construction of FQH condensates and quasiparticles in uniform magnetic field. (a) Illustration of a 2/5 island surrounded by the 1/3 condensate. The size of the rectangles reflects both: electronic charge density (vertical) and the mean 2D area occupied by the CF unit (horizontal). (b) Laughlin quasiparticles exist only in parent FQH condensates, illustrated by the "enclosed by" condensate unit. (c) Illustration of excitation of two $e/5$ quasiholes upon addition of flux $h/e$ to a 2/5 condensate island. However: a quasiparticle is a collective excitation of many condensate blocks and does not have a well-defined 2D area. The fractional charge of the quantized quasiparticle excitation is the total integrated charge deficiency or excess thus created in the parent condensate. Thus, these illustrations do not capture the FQH physics completely.

The primary insights here are: (a) The CF quasiparticle excitations of a FQH condensate are *not* simple additions or subtractions of a CF locally, but instead involve a collective excitation of the condensate with the total CF configuration differing by one CF or one vortex. A 2D electron system is comprised of an incompressible CF condensate and any CF quasiparticles. This parallels the excitation of a Laughlin quasiparticle, where condensate still exists after excitation, so that the topological order and incompressibility of the condensate is preserved when quasiparticles are present in the ground state. Because of the one-to-one correspondence between the CF model and the Haldane-Halperin construction, all microscopic processes and the resulting conclusions must be identical. (b) When considering CF excitations, it is essential to include the parent condensate wherever different filling FQH condensates coexist in a fixed uniform $B$. In such a system, the corresponding CF condensates have unequal $B^{CF} = B - 2jnh/e$, so that the attached flux and the Landau quantization area $S_0 = 2\pi\ell_0^2$ accounting becomes awkward. (c) It is not correct to assign a specific area to an excited CF quasiparticle. For example, excitation of quasiparticles out of condensate can be accomplished by variation of $B$; any given area contains correspondingly varying flux, and excitation of quasiparticles involves transfer of CFs between the condensate and the quasiparticles *within* that area. Excitation of a CF quasiparticle out of condensate does not involve any CF flux or charge transfer *outside* of the vicinity where the quasiparticle is created, it merely changes the microscopic CF configuration near the location of that quasiparticle. The intrinsic topological order and incompressibility of FQH condensates can thus be encoded in their composite fermion model structure.

As a CF model of the quasiparticle interferometer, consider a 2/5 condensate island of $N'$ 2-vortex CFs occupying area $S$ embedded in the 1/3 condensate of $N \gg N'$ CFs, a total of $N + N'$ electrons in the 2D plane. The unprojected wave function is

$$\Psi^{CF}_{2/5 \text{ in } 1/3} = D^2 \Psi_{2,N';1,N}, \qquad (31)$$

with the Slater determinant containing two filled Landau levels with $N'$ electrons up to orbital with $m = N'/2 - 1$, from $m = N'/2$ to $m = N'/2 + N - 1$ only the lowest Landau level is occupied:

$$\Psi_{2,N';1,N} = \begin{Vmatrix} \psi_0^0(z_1) & \psi_0^0(z_2) & \cdots & \psi_0^0(z_{N'+N}) \\ \vdots & \vdots & \cdots & \vdots \\ \psi_{N'/2+N-1}^0(z_1) & \psi_{N'/2+N-1}^0(z_2) & \cdots & \psi_{N'/2+N-1}^0(z_{N'+N}) \\ \psi_0^1(z_1) & \psi_0^1(z_2) & \cdots & \psi_0^1(z_{N'+N}) \\ \vdots & \vdots & \cdots & \vdots \\ \psi_{N'/2-1}^1(z_1) & \psi_{N'/2-1}^1(z_2) & \cdots & \psi_{N'/2-1}^1(z_{N'+N}) \\ 1 & 1 & \cdots & 1 \\ \vdots & \vdots & \cdots & \vdots \end{Vmatrix}. \qquad (32)$$

The static filling factor variation is achieved by stepping the electron density (thus defining the island boundary) in a uniform magnetic field $B$. An appropriate positive charged background maintains the charge neutrality of the total system. Increasing magnetic field to $B'$, and therefore the flux through $S$, excites $e/5$ quasiholes from the 2/5 CF condensate, whose electron density $\rho_{2/5} = -2e/5S_0$ increases in proportion, so that the total island charge and $N'$ do not change, as discussed in Sects. II and III above. By the literal interpretation of the CF construction, the topological order of the 2/5 FQH condensate implies that the 2/5 island must contain an integer multiple of the 2/5 CF condensate blocks, even in the presence of $e/5$ quasiparticles. This is equivalent to the island containing an integer number of MDD quasielectrons in the Haldane-Halperin hierarchy, Sect. III. Thus, the minimal periodic island reconstruction comprises addition of one more 2/5 CF block whose two CFs and one vertex come from the ten excited $e/5$ quasiholes: $-2e + 10(e/5) = 0$ and $5(h/e) - 10(h/2e) = 0$.

In parallel to the island reconstruction sub-periodic steps model presented in Sect. III, excitation of one $+e/5$ quasihole adds charge $-e/5$ and flux $h/2e$ to the 2/5 condensate, which can not be incorporated into the existing condensate CF blocks, so that the 2/5 condensate shrinks by $\tfrac{1}{2}S_0$. A polarization-charged annulus of $+e/30$ develops, which costs (small) Coulomb energy. After several of such steps, it becomes energetically favorable to incorporate one more 2/5 CF condensate unit into the island rather than to continue increasing the charging energy, with the charged annulus reversing its charge polarity. The period is completed when ten $+e/5$ quasiholes add charge $-2e$ and flux $5h/e$ to the 2/5 condensate, which are used to build the new 2/5 CF block. Upon this period, the 2/5 condensate reverts to its original equilibrium area and no net-charged annulus exists.

As shown in Fig. 2(b), an $-e/3$ quasielectron block of the 1/3 FQH fluid has the same CF content as a block of the $f=2/5$ condensate. Thus, quantization of the $-e/3$ quasielectron path encircling an incompressible 2/5 condensate can be stated in the CF language as the requirement for the encircling paths to contain an integer multiple of the 2/5 CF condensate blocks. The superperiod still comes from addition of one more island-encircling $-e/3$ quasielectron state. In contrast to the Haldane-Halperin hierarchy construction, in the CF model it is not evident that the additional $-e/3$ quasielectron state, $M \to M + \Delta_M$, is implied by the additional 2/5 CF condensate block without postulating the minimal 2/5 condensate two-CF building block, as shown in Fig. 2(a).

## VII. UN-PHYSICAL EXAMPLES

Without the constraint of the fundamental FQH physics, one can imagine numerous quasiparticle, CF, or "flux quanta" transfer processes. As an example, consider addition of flux $h/2e$ to the island. Using the Laughlin's gedanken experiment as analogy, this shrinks the island condensate by ½$S_0$. Assuming the island condensate can have arbitrary area, exciting one $e/5$ quasihole out of the condensate restores the island to the original area, the island remains neutral. Note that no $e/3$ quasiparticles are involved. Thus, neglecting the symmetry properties of the FQH condensates, the predicted island reconstruction periodicity is one island quasiparticle, same as in quantum antidots.[5,20,23] The island FQH condensate, however, can not have an arbitrary area, as shown in Sect. III.

Increasing magnetic field to $B' = B + h/eS$, the number of new $S'_0$ fitting into the unchanged island $S$ ("number of flux quanta") increases by one. Using the CF quasiparticle building blocks in Fig. 2(c), this breaks up one five-$S'_0$ condensate blocks into two three-$S'_0$, $e/5$ quasihole blocks. Thus, the island area $S$ is unchanged, $5+1=2\times 3$. This may seem to imply an $h/e$ island reconstruction period. However, as discussed above, such small CF building blocks shown in Fig. 2 do not capture the essential FQH physics. The quasiparticles are collective excitations and can not be represented by a single-CF or vortex transfer process. In order to capture the essential FQH physics, quasiparticle-containing FQH fluid must still be considered as comprised of an *incompressible* CF condensate plus its charged collective excitations, the quasiparticles. In an island geometry, the total island FQH fluid contains an integer multiple of CF condensate blocks. The individual CFs for quasielectrons (or their absence for quasiholes) are thus "shared" between all the condensate blocks of the fluid, contributing equally to their charge and flux (area).

Also, consider transfer of one CF from the interior of the 2/5 island to the outside boundary (Ohmic contacts), that is, excitation of one $e/5$ island quasihole as in Fig. 2(b); the resulting one-CF block is incorporated into the 1/3 condensate. If magnetic field were kept fixed, the island filling is still exactly 2/5, the island charges by $-e$ and shrinks by $5S_0$ to $S' = S - 5S_0$, where $S_0 = h/eB$, so that the island electronic charge density is now greater than the neutralizing background's. Also, the 2/5 island is now surrounded by an annulus of net positive charge comprised of 1/3 condensate in the higher-density positive neutralizing background. Therefore, such a process cannot lead to a periodic behavior because it leads to systematic charging of the island and thus huge Coulomb energy after ~10 such events.[8,9]

As another CF-counting example, consider subtraction of a CF from the 2/5 condensate (creation of an $e/5$ quasihole as in Fig. 2(b), block area $3S_0$), and its addition to the 1/3 condensate (creation of an $-e/3$ quasielectron, block area $5S_0$). Creation of five $e/5$ quasiholes in the 2/5 condensate, total charge deficiency $e$, shrinks the embedding 2/5 condensate area by $10S_0$. This process can not be represented as an equivalent excitation of three $-e/3$ quasielectrons from the 1/3 condensate, total excess charge $-e$, which expands the embedding 1/3 condensate area by $6S_0$. A charge transfer process, annihilation of the five $e/5$ quasiholes with the three $-e/3$ quasielectrons, shrinks the 2/5 island (and extends the 1/3 condensate) by $4S_0$. Such a process is a gedanken shrinkage of the island area while maintaining fillings fixed everywhere, thus it corresponds to neither changing uniform magnetic field nor to any physical gate action. In addition, a net charged disc-annulus dipole is created leading to macroscopic Coulomb energy, as above. From these examples we see that, in general, either the charge is not conserved, or the areas (in units of $S_0$, that is flux) do not match. In any case, such CF-counting exercises fail to incorporate essential quantum Hall physics.

## VIII. CONCLUSIONS

To conclude, two points are in order. (A) We note that a consistent theory of the fractionally charged elementary excitations of a FQH fluid (Laughlin quasiparticles) must incorporate their anyonic statistics. This is evident even in the quantum antidot geometry, where a straightforward description of the $M \to M + \Delta_M$ sequential transitions is possible only explicitly including the anyonic statistical contribution,[20] whereas an integer quasiparticle statistics requires a *different* singular gauge for each $M$. Periods $\Delta_\Phi = h/e$ and $\Delta_Q = e/3$, the same as in quantum antidots, have also been observed in a Laughlin quasiparticle interferometer containing 1/3 fluid only.[34] It is not possible to "gauge away" the anyonic term in Eq. (24), describing the superperiodic oscillations observed in the 2/5 in 1/3 quasiparticle interferometer,[8,9] because the interior of the interference path contains 2D electrons everywhere.

(B) While it is possible to construct a phenomenological composite fermion model of FQH condensates without explicit reference to statistics of quasiparticles, proper composite fermion representation of Laughlin quasiparticles is fully equivalent to that in the Haldane-Halperin hierarchical construction. Accordingly, composite fermion illustration of Laughlin quasiparticles does require quasiparticle anyonic statistics in order to be self-consistent and to agree with the experiment. For example, without regard to the constraints imposed by the topological order and incompressibility of the condensates, there is no reason why the experimental period would not correspond to excitation of a single $e/5$ quasiparticle out of the 2/5 condensate. Even if the island were charged and the Coulomb blockade were present, it still would be possible to charge the island in increments of $e = 5(e/5) = 3(e/3)$, contrary to the experiment. The underlying topological order of the FQH condensates is thus demonstrated through the observable anyonic statistics of their elementary charged excitations.

**Acknowledgements**

Stimulating discussions with B. I. Halperin and D. V. Averin are gratefully acknowledged. This work was supported in part by the National Science Foundation and by the US Army Research Office.